%% file: ms.tex
\documentclass[12pt,preprint]{aastex}

\shorttitle{The Metal Line Variability of G29-38} 
\shortauthors{Debes \& L\'opez-Morales}

\begin{document}
\title{A Second Look at the Metal Line Variability of G29-38\altaffilmark{1}}
\author{John H. Debes\altaffilmark{2} \& Mercedes L\'opez-Morales\altaffilmark{2,3}}

\altaffiltext{1}{Based on data gathered with the 6.5 meter Magellan Telescopes located at Las Campanas Observatory, Chile}
\altaffiltext{2}{Carnegie Institution of Washington, Department of Terrestrial
Magnetism, \\ 5241 Broad Branch Rd. NW, Washington D.C., 20015, USA}
\altaffiltext{3}{Hubble Fellow}

\begin{abstract}

The pulsating white dwarf G29-38 possesses a dust disk and metal lines
attributed to the accretion of its disk material. \citet{vonhipg29} have reported variability in the equivalent width of G29-38's CaII K line on the timescale of days. We use high resolution optical spectroscopy of G29-38's CaII K line to test this observation.
Over six days spanning in June 2007 and October 2007 we see no evidence for variability in the equivalent width of the Ca II K line.  We also sample the variability of the Ca II K line over integrated timescales of $\sim$100-500 seconds, where errors from incomplete coverage of pulsation modes are predicted to be $\sim$8-15\%.   We find that the scatter of the equivalent widths over this time period is consistent with measurement errors at the 7\% level, slightly weaker than predicted but within the uncertainties of predictions.  Weaker Ca and Mg lines observed show no significant variability on yearly timescales over ten years based on our data and other high resolution
spectra.  We conclude that further study is warranted to verify if the accretion onto G29-38 is variable.
\end{abstract}


\section{Introduction}
G29-38 is a fascinating white dwarf.  It is at a distance of 14~pc \citep{vanaltena95}, it pulsates \citep{kleinman98}, it shows an atmosphere polluted with Ca, Mg, and Fe \citep{koester97}, 
and it posseses an infrared excess attributed to a dust disk within its tidal
 disruption radius \citep{zuckerman87,jura03,reach05b}.  Since tidally disrupted planetesimals are a likely explanation for the dust disk and G29-38's metal pollution, it is
 tempting to speculate on the presence of a planet or planets
 perturbing 
objects close to the white dwarf \citep{debes02}.  This perturbation, which could be constant, periodic, or stochastic, may cause variability in the amount
of material that drizzles onto the atmosphere of G29-38 from the disk.  

Several factors complicate the search for accretion variability in G29-38.  
G29-38 is a ZZ Ceti pulsator at the cool end of the DA variability strip, with peak-to-peak continuum 
changes of a few percent \citep{winget90,kleinman98}.  These pulsations change over monthly to yearly timescales, but the dominant modes tend to be around 600~s.
Spectroscopic determination of its fundamental atmospheric parameters, i.e. 
T$_{eff}$ and log $g$\ (cgs), are uncertain due to the pulsations. 
For example, \citep{koester06} 
find $T_{eff}$ = 12,900 K and log $g$\ = 7.9, while \citep{bergeron04} find $T_{eff}$ = 11,820 K and log $g$\ = 8.14. Accurate 
determination of the temperature and gravity is important 
in the case of G29-38, since diffusion calculations show orders of magnitude different settling times between 9000~K and 14000~K (see Figure 2 in \citet{koester06}).  With the two different measurements, G29-38's predicted settling timescale would vary between 4 and 7 days.  However, further uncertainties may be present for pulsators.
Empirical observations of equivalent width variations can directly probe diffusion calculations 

\citet[][hereafter vHT07]{vonhipg29} investigated the equivalent width variability of the strong CaII K line by compiling all spectroscopic observations of G29-38 from 1996-2000.  This data included low resolution datasets used to monitor variability in the line-of-sight velocities and line profiles of hydrogen lines arising from G29-38's pulsations \citep{vankerk,thompson06}.  This variability is correlated with pulsations of the white dwarf and are shorter than the settling time for metals \citep{koester06}. vHT07 found a significant dip in the equivalent width of the CaII K line of G29-38 between 
1997 and 1999, from averages of the two low-resolution datasets. The measured equivalent width went from 167$\pm4$~m\AA~ to 280$\pm$8~m\AA.  They attributed this dip to a change in accretion rate, but vHT07 were cautious about the effect of the pulsations on variability of the Ca line strength.  They 
used an analysis 
of their time series spectra to estimate the effect incomplete coverage pulsations may have on theline strength.  They 
estimated that for observations less than an hour, errors due to pulsations would be less than 5\%.  However, shorter exposure times were more sensitive to incomplete pulsation coverage; with errors ranging from 8\% for a ten minute exposure to a 13\%
error for a five minute exposure were possible. 
Based on their time series analysis, they concluded that the CaII K line strength changes they observed were not due to pulsations.  vHT07 subsequently combined the two low resolution datasets with several other high resolution spectroscopic observations spanning between 1996 and 2000 and predicted that the CaII K line must be variable on timescales as short as fifteen days.

To test for further variability, we observed G29-38 in two multi-day runs separated by four months for a total of 19 individual observations with exposure times ranging from 100 to 500 seconds.  All individual observations are short enough to determine the extent of possible systematic errors introduced by pulsations.  When averaged together, each run covers enough time ($\sim$40 minutes) to average over potential pulsations and get an idea of the true average strength of the line.  We therefore expected that our data in aggregate would be accurate to less than a few percent, but we expected that any single observation might vary by as much as 7-13\%.  The observations, data analysis, and our conclusions are detailed in the following sections of this paper.

\section{Observations}
\label{sec:obs}
We observed G29-38 with the blue chip of the MIKE spectrograph \citep{bernstein03} installed at the 6.5-m Magellan Clay Telescope at Las Campanas Observatory (Chile). Six spectra were collected on UT 2007 June 25--26 and another 13 spectra on UT 2007 October 24-27. Both runs used a 0\farcs7$\times$5\farcs0 slit, yielding an average spectral resolution of R = 40000 at 3933 \AA. The spectra cover wavelengths between 3335 and 5120 \AA.   Typically three observations were taken per night with the exception of 24 October, where one 100s exposure was taken and then three 200s exposures.  The exposure times in June were determined  by poor (1\arcsec-1\farcs3) seeing and were between 300 and 500 seconds, whereas spectacular ($\sim$0\farcs5) seeing conditions were present for the first two nights of the October run, followed by nominal seeing ($\sim$0\farcs8) the last two nights, allowing shorter exposure times.  
The signal-to-noise of each exposure is roughly 25, for each night is roughly 50, and for the averages between runs is about 80.
 
\begin{figure}
\plotone{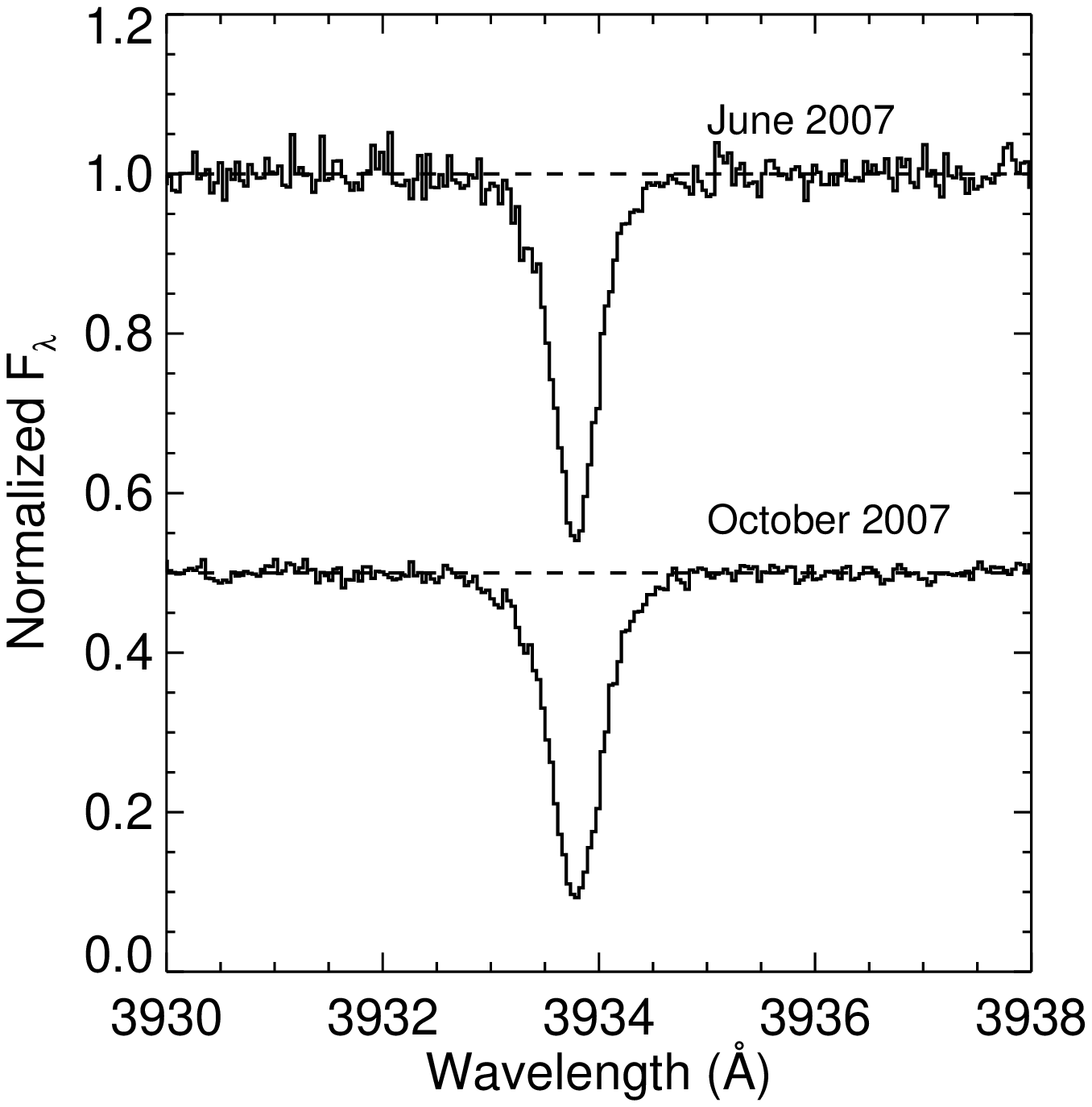}
\caption{\label{fig:spec}Comparison of the CaII K line in G29-38 from the combined spectra between the June 2007 and October 2007 Magellan runs.}
\end{figure}

The data were extracted and flatfielded using the MIKE reduction pipeline written by D. Kelson, with methodology
described in \citet{kelson00}, \citet{kelson03}, and \citet{kelson07}.  All six individual spectra from the June run, 2400 seconds in total, and all thirteen individual spectra from the October run, 2500 seconds in total, were combined into averaged spectra for each run to search for weak lines other than Ca.  An example of the fidelity of these combined spectra is presented in Figure \ref{fig:spec}.

Absorption lines from Mg II, Ca I, and Ca II were detected.  In particular, the Ca II K line and the Mg II $\lambda$4481 line fell on two different orders, a feature that we took advantage of to minimize systematic errors
 in the determination of the equivalent widths.  Other lines from Mg I were detected at 3830, 3833, and 3838\AA.  Ca I was detected at 4227\AA, and the Ca II H line and another Ca II line at 3737\AA\ were similarly detected.

In order to claim variability in a measured equivalent width, a careful examination of the measurement procedure and an accounting of systematic errors must be done. This step is important to robustly compare observations done at different telescopes (see \S \ref{sec:var}).  To this end we developed a method for measuring the equivalent width of a line while ensuring that all the line optical depth was measured and that all systematic errors were accounted for.  Inspection of the average spectra from June and October show that the line profile differs between the two nights and that the lines deviate from pure Gaussian shapes (See Fig \ref{fig:spec}).  The variable profile and slightly asymmetric wings may be caused by shifting velocity fields on the WD surface \citep{koester07}.  We  decided to directly integrate the line profile from the data, rather than
fitting the line with a Gaussian and missing contributions to th equivalent width.  

The continuum of some lines is affected by curvature introduced by the proximity of strong hydrogen Balmer lines. That curvature is removed by fitting a high order polynomial.  
The choice of how to fit the continuum and the size of the continuum window to use can introduce systematic errors in the equivalent width.  We used a polynomial fit to the continuum with a window of 
at least $\pm$5\AA~ on each side of the line center. To account for the systematic errors, we measured the equivalent width over all combinations of window size from $\pm$5\AA~ to $\pm$10\AA~ and polynomial fits ranging from third to twelfth order.  We took the median measure of the equivalent width and the sum in quadrature of the measured standard deviation of measurements (the systematic error) and the estimated error due to signal-to-noise.

An appropriate window around a line must be chosen to accurately measure the total equivalent width.  We determined an optimal window by
 measuring the equivalent width for the averaged spectra from each run using a range of window sizes.  Once the continuum is fitted, the equivalent width will peak at a radial distance from the line center where the line no longer contributes.  In the particular case of the CaII K line, we determined this distance to be $\pm$1.2--1.5~\AA.  

In addition to the Ca II K line, we measured the equivalent width for some of the weaker lines we detected, Mg II $\lambda$4481, Ca I $\lambda$4227, and the Ca II H line.  For these lines, we restricted ourselves to the averages of the two runs.

\section{Testing Metal Line Variability}
\label{sec:var}
\begin{figure}
\plotone{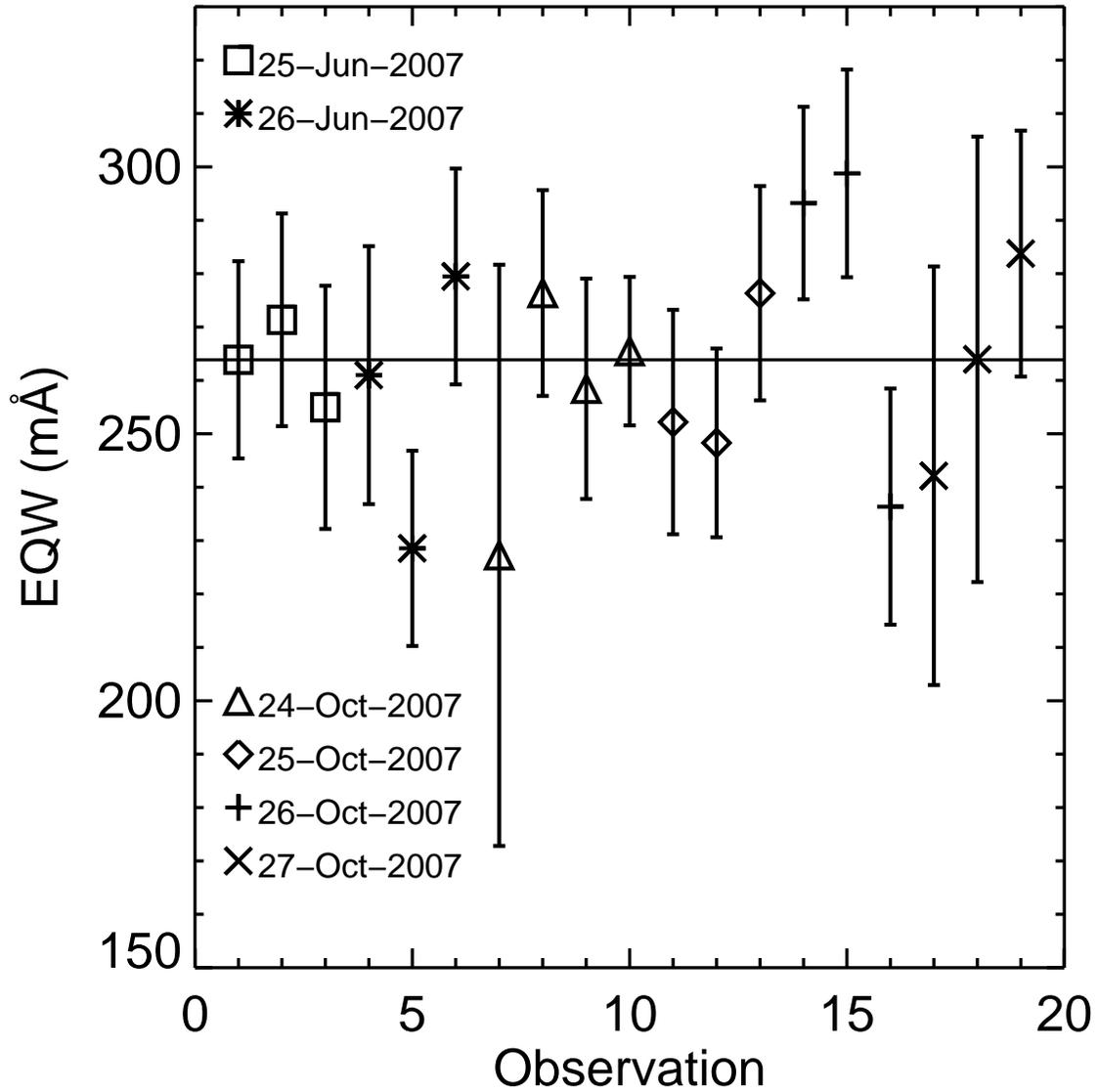}
\caption{\label{fig:eqw} Equivalent width of G29-38's CaII K line from each observation in our June and October 2007 runs.  The first six points are from theJune run and exposure times range between 300 and 500 sec in duration, while the last 13 points are from the October run. Exposure times in this case range betweem 100 and 200 seconds. The median of all the observations,  264$\pm$20m\AA, is represented by the horizontal solid line.}
\end{figure}

Figure \ref{fig:eqw} shows the equivalent width of the CaII K line from each exposure, in order of ascending sequential observation, i.e. the first spectrum taken in June is observation 1 and the last spectrum in October is observation 19.  Each night has a specific symbol on the plot.  

We find that the equivalent width of the CaII K line is stable over the time period we observed G29-38.  The standard deviation of all 19 measurements is 20 m\AA, while the median estimated error of the measurements is 20 m\AA.  These results suggest that if there is variability due to pulsations for these spectra it is $<$ 7\%.  This is less than
a factor of two lower than that predicted by vHT07.  We can then
 add observations over a small number of pulsational periods to increase signal-to-noise without fear of being subject to systematic errors.  

Averaging the two runs gives equivalent widths of 262$\pm$9~m\AA~
and 269$\pm$7~m\AA~ for the June and October runs, respectively.  These values
are thus consistent with
no variation in line strength over 4 months at the 2\% level.  If short term variability occurred between the two measurements we would not have detected it.
This implies that the underlying source of accretion is steady.  
Table \ref{tab:tab} shows the results for our June and October runs (MJD 54278.5 and 54399.1, respectively) for all species measured.

It is relevant to revisit whether there has been significant variation of the equivalent width of the CaII K line or any of the other lines since 1997,
the dates of the last published values.  There exist eight other measurements, mentioned in vHT07, including results from \citet{koester97},\citet{zuckerman03}, and \citet{koester06}.  Excluding the low resolution data from vHT07, 
we are left with six high resolution spectra of G29-38 in addition to our own data.  Of the six, three are from the Keck Observatory and are publicly available \citep[see][]{zuckerman03}, while two spectra from the ESO VLT were kindly provided by D. Koester.  We disregard the discovery equivalent width due to its 
large error.  

\begin{figure}[h]
\plotone{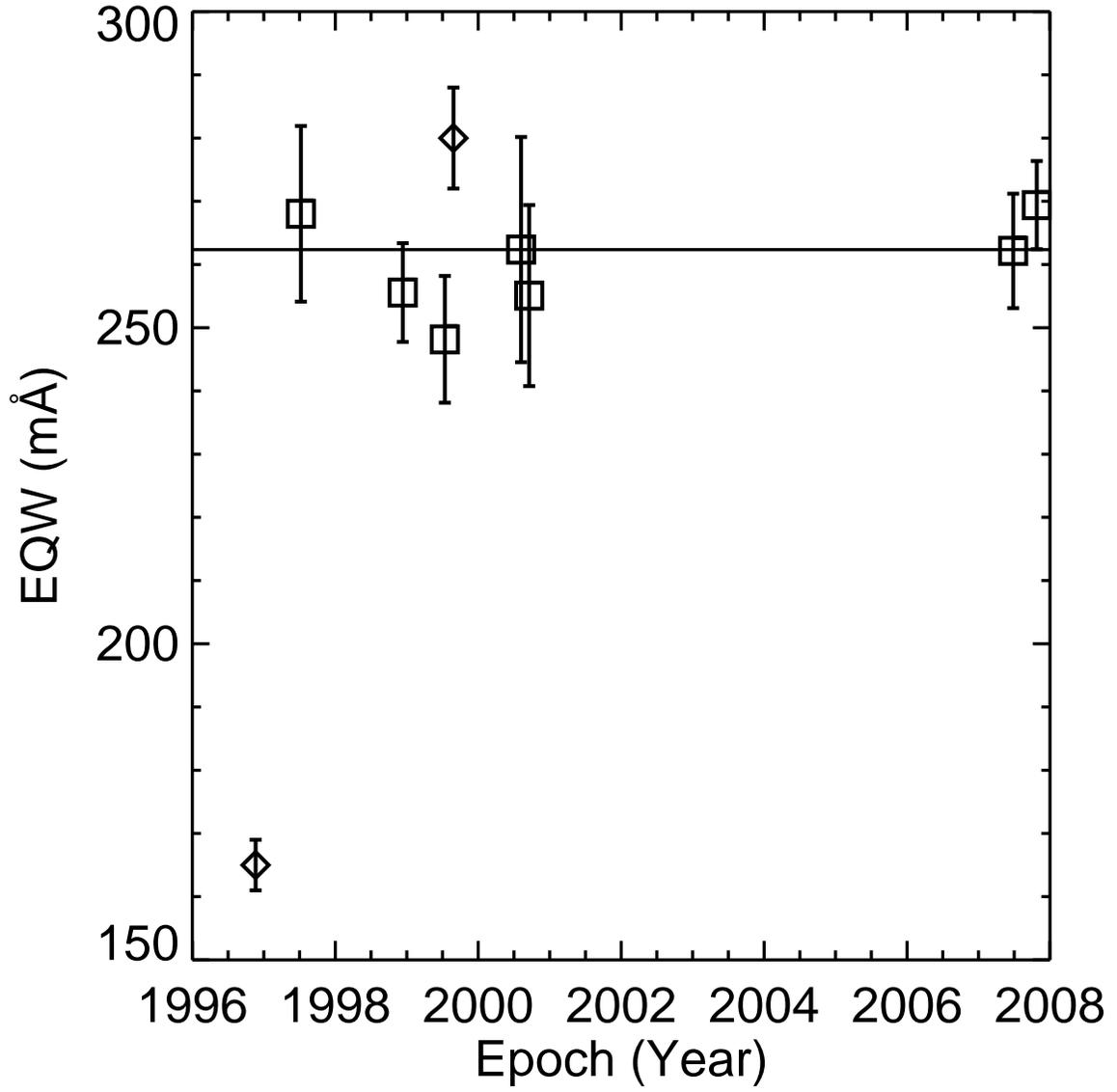}
\caption{\label{fig:final} Equivalent width of G29-38's CaII K line spanning the past decade.  Squares are from high resolution data, while diamonds are from
low resolution data. The horizontal line indicates the median of all the observations, not including the low resolution data reported in vHT07.}
\end{figure}

We re-measured the equivalent width of the CaII K line using the same method described in \S\ref{sec:obs}, and compared the new values to our data.  Figure \ref{fig:final} shows the results, including the low resolution data reported in vHT07.  Neglecting the two low resolution points, the median of the seven high resolution observations is 262~m\AA, with a standard deviation of 7.6~m\AA~ and a median measurement error of 10.6~m\AA.  With the exception of the low resolution points, the underlying source of accretion for G29-38 is incredibly stable over a timescale of 10 years.  If we add the
reported measurements of the low resolution data, the 1997 point remains significantly discrepant.  We compare the 1999.65 point with the median and find that it is consistent at $<$ 2$\sigma$.  However, as noted in \S\ref{sec:conc}, our measured equivalent widths are higher on average than vHT07's and so the 1999.65 point may also be discrepant if we were to remeasure that data.

\begin{figure}[h]
\plotone{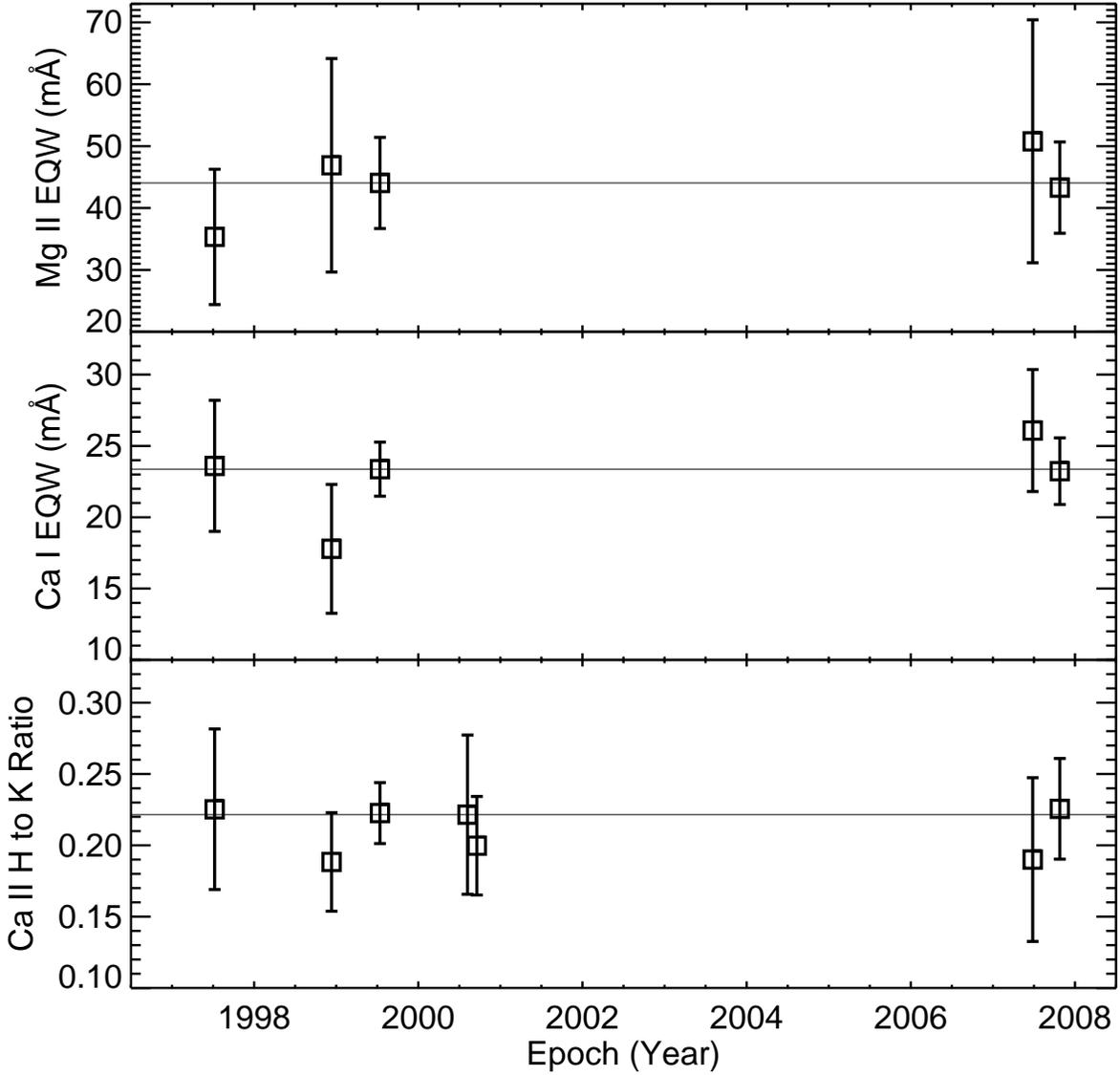}
\caption{\label{fig:other} (Top) Plot of measured Mg II $\lambda$4481 line equivalent width, horizontal line is median value of 44$\pm$6~m\AA.  (Middle) Plot of measured Ca I $\lambda$4427 line equivalent width,horizontal line is median value of 23$\pm$3~m\AA.  (Bottom) Plot of measured CaII H to K equivalent width ratios for G29-38, horizontal line is median value of 0.22$\pm$0.02}
\end{figure}

We also measured the equivalent widths of the other three metal lines in the Keck and VLT data. The results, combined with our data, are shown in Figure \ref{fig:other}. Similar to the strong Ca II K line, these weaker lines 
detected in G29-38 show no variability.  The Mg II $\lambda$4481 line is visible in our data as well as the Keck data, but it is not reliably detected in the VLT data due to insufficient resolution.  Over the five available observations 
this line is stable at a 12\% level, with a median equivalent width
of 44$\pm$6~m\AA.  The Ca I $\lambda$4427 line was also detected in our data and the Keck data, but was not detected in the VLT data for the same reason as the MgI line above.  Over the five available observations, the Ca I line equivalent width is 23$\pm$3~m\AA, and is stable at a 13\% level.  Finally, we measured the Ca II H line, which resides in the wings of the H$_\epsilon$ line. This line also has a stable equivalent width of 55$\pm$5~m\AA over ten years at a 10\% level. This line was detected in all the availale data.  The results can be found in Table \ref{tab:tab}

Given optically thin conditions for the CaII H and K lines, their strength should be in the ratio of H to K of 1:2.  We plot the observed line ratios over the Keck, VLT, and Magellan data in the bottom diagram in Figure \ref{fig:other} ( The ratio is 22$\pm$2\%, which can be explained by additional opacity at the H line due to the H$_\epsilon$ line wings (D. Koester 2007, personal communication).

\section{Conclusions}
\label{sec:conc}

We have shown that G29-38's Ca II K line has not significantly varied recently, or over the past decade, with the exception of one or possibly
two observations.  A stable result is also obtained for the other metal lines Mg II $\lambda$4481, Ca I $\lambda$4427, and CaII H.  We see no evidence for
systematic errors due to pulsations at around the 7\% level in our observations, slightly at odds with the predictions of vHT07.  However, G29-38's modes are variable in both period and strength, while vHT07 calculated errors for a worst case scenario involving strong pulsations.  The last measurements of G29-38's pulsations that we are aware of occurred in 2005 and the strongest modes had flux variations of $\sim$1\%, whereas the flux variations reported for vHT07's low resolution data were $\sim$3\% (M. Montgomery 2008, private communication).  It is therefore not surprising that we don't see evidence for 
errors due to pulsations.  Other calculations of equivalent width variability due to pulsations also predict errors at the level of 10\%, suggesting that when accounted for, these errors have minimal impact on long term observations \citep{koester07}.

One point to note is the different values that we derive in our measurement of the archival equivalent widths when compared to those obtained by vHT07. Those differences come from the different techniques used in the mesurement of the equivalent width and in the continuum fitting, which highlights the need for a consistent method when comparing observations from different sources.  The fact that our method produces on average larger equivalent widths than those reported by vHT07 can be related to incorrect accounting for line flux in the wide wings of the CaII K line profile when using Gaussian fits.

Since the low resolution data are the only points that deviate more than a few percent from the median equivalent width value, it is important to determine whether low resolution observations could give spurious results at levels greater than their measured errors.   The lack of any flatfielding in vHT07, as well as changes in the wings of the $H_{\epsilon}$ line may all affect continuum fitting, something that low resolution data will be more sensitive to.  To test this hypothesis, we rebinned our data down to different resolutions that approach the 7-8\AA level
used in vHT07 to see if we could recover the same equivalent width using our
method outlined above.  At this level of rebinning, our S/N exceeded that reported in vHT07.  We used a window for measuring the equivalent width of 10\AA centered on the line and fit the continuum with 10\AA~ on either side of the window.  We  
tried resolutions that ranged from 0.04~\AA to 4~\AA.  We find for the average spectrum of our October data that the measured equivalent widths at the lowest resolutions deviated significantly from our measured value at high resolution, going as low as 140~m\AA~ to as high as 350~m\AA.  At resolutions less than 0.1~\AA, our results converge to within 1$\sigma$ of our previously measured value.  It is impossible to tell if vHT07's data suffers from a similar problem, but high resolution, high S/N confirmation of this variability is desired.

Whether the accretion onto G29-38 varies or not, the idea of variable accretion onto white dwarfs is an interesting line of study.  Spectroscopic monitoring of DAZs with close companions and short settling times offers an interesting insight into the stellar wind behavior of M dwarfs \citep{debes06b}, while variability in single DAZs may conceivably determine the source of material that pollutes the surface of these white dwarfs.  Both types of polluted white dwarfs directly test models of white dwarf diffusion times.  Few DAZs have been monitored in a systematic way, but the MIKE spectrograph provides a sensitive and stable platform with which to do long term studies.

\acknowledgments{MLM acknowledges support provided by NASA 
through Hubble Fellowship grant HF-01210.01-A awarded
by the Space Telescope Science Institute, which is operated by the
Association of Universities for Research in Astronomy, Inc., for NASA,
under contract NAS5-26555.}

\bibliography{g29bib}
\bibliographystyle{apj}

\clearpage
\input{Table1.tex}

\clearpage

\end{document}

%% file: Table1.tex
\begin{deluxetable}{ccccc}
\tabletypesize{\footnotesize}
\tablecolumns{5}
\tablewidth{0pc}
\tablecaption{\label{tab:tab} Equivalent Width Measurements}
\tablehead{ \colhead{MJD} & \colhead{Ca II K} & {Ca II H} & {Ca I} & {Mg I} \\
 \colhead{} & \colhead{(m\AA)} & \colhead{(m\AA)} & \colhead{(m\AA)} & \colhead{(m\AA)} }
\startdata
       50636.577 & 268$\pm$15 & 60$\pm$16 & 24$\pm$5 & 35$\pm$11 \\
       51158.200 & 256$\pm$9 & 48$\pm$10 & 18$\pm$5 & 47$\pm$17 \\
       51403.463 & 248$\pm$11 & 55$\pm$5 & 23$\pm$2 & 44$\pm$7 \\
       51762.681 & 232$\pm$12 & 57$\pm$13 & & \\
       51804.670 & 239$\pm$7 & 51$\pm$6 & & \\
       54278.500 & 262$\pm$9 & 50$\pm$15 & 26$\pm$4 & 51$\pm$20 \\
       54399.100 & 269 $\pm$7 & 61$\pm$9 & 23$\pm$2 & 43$\pm$7 \\
\enddata       
\end{deluxetable}

%% file: ms.bbl
\begin{thebibliography}{20}
\expandafter\ifx\csname natexlab\endcsname\relax\def\natexlab#1{#1}\fi

\bibitem[{{Bergeron} {et~al.}(2004){Bergeron}, {Fontaine}, {Bill{\`e}res},
  {Boudreault}, \& {Green}}]{bergeron04}
{Bergeron}, P., {Fontaine}, G., {Bill{\`e}res}, M., {Boudreault}, S., \&
  {Green}, E.~M. 2004, \apj, 600, 404

\bibitem[{{Bernstein} {et~al.}(2003){Bernstein}, {Shectman}, {Gunnels},
  {Mochnacki}, \& {Athey}}]{bernstein03}
{Bernstein}, R., {Shectman}, S.~A., {Gunnels}, S.~M., {Mochnacki}, S., \&
  {Athey}, A.~E. 2003, in Presented at the Society of Photo-Optical
  Instrumentation Engineers (SPIE) Conference, Vol. 4841, Instrument Design and
  Performance for Optical/Infrared Ground-based Telescopes. Edited by Iye,
  Masanori; Moorwood, Alan F. M. Proceedings of the SPIE, Volume 4841, pp.
  1694-1704 (2003)., ed. M.~{Iye} \& A.~F.~M. {Moorwood}, 1694--1704

\bibitem[{{Debes}(2006)}]{debes06b}
{Debes}, J.~H. 2006, \apj, 652, 636

\bibitem[{{Debes} \& {Sigurdsson}(2002)}]{debes02}
{Debes}, J.~H. \& {Sigurdsson}, S. 2002, \apj, 572, 556

\bibitem[{{Jura}(2003)}]{jura03}
{Jura}, M. 2003, \apj, 584, L91

\bibitem[{{Kelson}(2003)}]{kelson03}
{Kelson}, D.~D. 2003, \pasp, 115, 688

\bibitem[{{Kelson}(2007)}]{kelson07}
---. 2007, \pasp, submitted

\bibitem[{{Kelson} {et~al.}(2000){Kelson}, {Illingworth}, {van Dokkum}, \&
  {Franx}}]{kelson00}
{Kelson}, D.~D., {Illingworth}, G.~D., {van Dokkum}, P.~G., \& {Franx}, M.
  2000, \apj, 531, 159

\bibitem[{{Kleinman}(1998)}]{kleinman98}
{Kleinman}, S.~J. e.~a. 1998, Astrophys. J., 495, 424

\bibitem[{{Koester} \& {Kompa}(2007)}]{koester07}
{Koester}, D. \& {Kompa}, E. 2007, \aap, 473, 239

\bibitem[{{Koester} {et~al.}(1997){Koester}, {Provencal}, \&
  {Shipman}}]{koester97}
{Koester}, D., {Provencal}, J., \& {Shipman}, H.~L. 1997, \aap, 320, L57

\bibitem[{{Koester} \& {Wilken}(2006)}]{koester06}
{Koester}, D. \& {Wilken}, D. 2006, \aap, 453, 1051

\bibitem[{{Reach} {et~al.}(2005){Reach}, {Kuchner}, {von Hippel}, {Burrows},
  {Mullally}, {Kilic}, \& {Winget}}]{reach05b}
{Reach}, W.~T., {Kuchner}, M.~J., {von Hippel}, T., {Burrows}, A., {Mullally},
  F., {Kilic}, M., \& {Winget}, D.~E. 2005, \apjl, 635, L161

\bibitem[{{Thompson}(2006)}]{thompson06}
{Thompson}, S.~E. 2006, in Astronomical Society of the Pacific Conference
  Series, Vol. 352, New Horizons in Astronomy: Frank N. Bash Symposium, ed.
  S.~J. {Kannappan}, S.~{Redfield}, J.~E. {Kessler-Silacci}, M.~{Landriau}, \&
  N.~{Drory}, 289--+

\bibitem[{{van Altena} {et~al.}(2001){van Altena}, {Lee}, \&
  {Hoffleit}}]{vanaltena95}
{van Altena}, W.~F., {Lee}, J.~T., \& {Hoffleit}, E.~D. 2001, VizieR Online
  Data Catalog, 1238, 0

\bibitem[{{van Kerkwijk} {et~al.}(2000){van Kerkwijk}, {Clemens}, \&
  {Wu}}]{vankerk}
{van Kerkwijk}, M.~H., {Clemens}, J.~C., \& {Wu}, Y. 2000, \mnras, 314, 209

\bibitem[{{von Hippel} \& {Thompson}(2007)}]{vonhipg29}
{von Hippel}, T. \& {Thompson}, S.~E. 2007, \apj, 661, 477

\bibitem[{{Winget} {et~al.}(1990){Winget}, {Nather}, {Clemens}, {Provencal},
  {Kleinman}, {Bradley}, {Wood}, {Claver}, {Robinson}, {Grauer}, {Hine},
  {Fontaine}, {Achilleos}, {Marar}, {Seetha}, {Ashoka}, {O'Donoghue}, {Warner},
  {Kurtz}, {Martinez}, {Vauclair}, {Chevreton}, {Kanaan}, {Kepler},
  {Augusteijn}, {van Paradijs}, {Hansen}, \& {Liebert}}]{winget90}
{Winget}, D.~E., {et~al.} 1990, \apj, 357, 630

\bibitem[{{Zuckerman} \& {Becklin}(1987)}]{zuckerman87}
{Zuckerman}, B. \& {Becklin}, E.~E. 1987, Nature, 330, 138

\bibitem[{{Zuckerman} {et~al.}(2003){Zuckerman}, {Koester}, {Reid}, \& {H{\"
  u}nsch}}]{zuckerman03}
{Zuckerman}, B., {Koester}, D., {Reid}, I.~N., \& {H{\" u}nsch}, M. 2003, \apj,
  596, 477

\end{thebibliography}
